\documentclass{article}

\usepackage{microtype}
\usepackage{graphicx}
\usepackage{subfigure}
\usepackage{booktabs}
\usepackage{caption}
\usepackage{subcaption}
\usepackage{wrapfig}
\usepackage{amsmath}
\usepackage{amssymb}
\usepackage{mathtools}
\usepackage{amsthm}
\usepackage{xcolor}
\usepackage{hyperref}
\usepackage[accepted]{icml2026}

\makeatletter
\renewcommand{\Notice@String}{\textit{Conference on Physics and AI at Stanford University (PAI 2026).}}
\makeatother

\theoremstyle{plain}

\theoremstyle{definition}

\theoremstyle{remark}

\icmltitlerunning{The Galaxy's Guide to the Tokenizer}

\begin{document}

\twocolumn[
\icmltitle{The Galaxy's Guide to the Tokenizer: \\
           A Benchmark for Scientific Foundation Models}

\begin{icmlauthorlist}
\icmlauthor{Sogol Sanjaripour}{ucr}
\icmlauthor{Michael J.\ Smith}{astroai,cfa,utbd}
\icmlauthor{Manuel P\'erez-Carrasco}{astroai,cfa}\\
\icmlauthor{Juan Rafael Mart\'inez-Galarza}{astroai,cfa}
\icmlauthor{Bahram Mobasher}{ucr}
\icmlauthor{Gabriela Canalizo}{ucr}
\end{icmlauthorlist}

\icmlaffiliation{ucr}{University of California, Riverside}
\icmlaffiliation{astroai}{AstroAI}
\icmlaffiliation{cfa}{Center for Astrophysics $\vert$ Harvard \& Smithsonian}
\icmlaffiliation{utbd}{UniverseTBD}

\icmlcorrespondingauthor{Sogol Sanjaripour}{sogol.sanjaripour@email.ucr.edu}

\icmlkeywords{Machine Learning, ICML, Foundation Models, Astronomy, Tokenization}

\vskip 0.3in
]

\printAffiliationsAndNotice{}

\begin{abstract}
Tokenization is central to adapting scientific data for transformer-based foundation models, yet its impact on learned representations remains poorly understood.
We compare four tokenization strategies---Affine, AIM, JetFormer, and VQ-VAE---within a unified transformer framework for astronomical imaging.
Using $640\,000$ galaxy images from the DESI Legacy Survey and a shared AstroPT backbone, we evaluate each method on reconstruction fidelity and prediction of physical properties.

Our results reveal trade-offs across approaches.
The flow-based JetFormer achieves higher reconstruction quality, while VQ-VAE yields strong probe performance for galaxy physical properties.
Affine and AIM better preserve localized morphological information.
We find that reconstruction and representation quality are decoupled, and no single method consistently performs best across the tasks considered here.
By grounding our evaluation in independently measured physical quantities, we hope this study serves to highlight the potential of scientific data as a basis for constructing interpretable benchmarks for foundation models.
\end{abstract}

\begin{figure*}[t]
    \centering
    \includegraphics[width=0.8\textwidth]{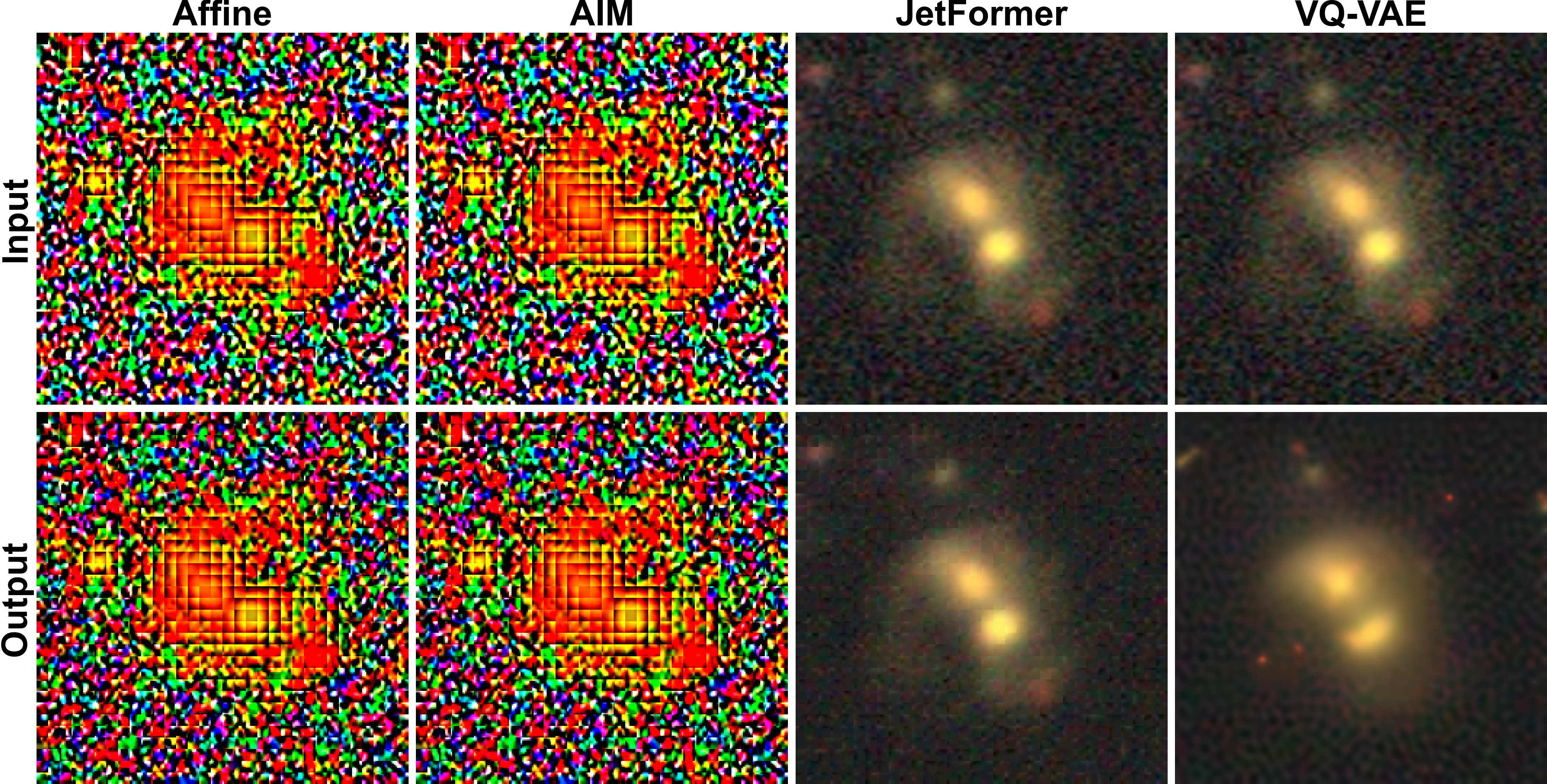}
    \caption{Comparison of image reconstruction quality using different tokenization strategies with a shared transformer backbone (AstroPT). The top row shows the input galaxy images, while the bottom row presents the reconstructions produced by each tokenizer. For the AIM and Affine models, both input and reconstructed images are shown in their patch-wise normalized representation.}
    \label{fig:tokenizer_reconstruction_comparison}
\end{figure*}

\section{From pixels to physics}

The Big Data Era of astronomy has arrived.
Surveys such as Euclid \citep{ref_euclid}, LSST \citep{ref_lsst}, and DESI \citep{ref_desi} are producing imaging data at unprecedented scale, and transformer-based foundation models have emerged as a leading paradigm for extracting scientific insight \citep[e.g.][]{ref_smith2023,ref_leung2023,ref_koblischke2024}.
Yet, while architectural choices have been widely studied, far less attention has been paid to \emph{tokenization}---the process by which observations are converted into sequences that transformers can process---and how it shapes learned representations.

Scientific domains provide a natural setting for this question, as observations are governed by known physical processes and accompanied by independently measured quantities that serve as objective ground truth.
In astronomy, these include continuous properties such as redshift, stellar mass, and star formation rate, enabling physically grounded evaluation of representations beyond task-specific metrics \citep{sanjaripour2024,sanjaripour2025,Hemmati2026}.
These questions also have practical relevance.
Astronomical foundation models increasingly aim to unify multiple modalities within a shared framework \citep[see e.g.][]{ref_mmu2024,ref_euclid2025,ref_parker2025}, and tokenization strategies that perform well for one modality may not generalize to others.
Moreover, representations optimized for reconstruction may discard information relevant for scientific inference, a concern also observed in machine learning \citep[e.g.,][]{ref_bfl_representation,ref_mousavi2025}.

We systematically compare four tokenization strategies within a controlled framework: Affine (linear) projection; MLP-based \citep[AIM-like;][]{ref_elnouby2024}; flow-based \citep[Jetformer-like;][]{ref_tschannen2024}; and VQ-VAE-based approaches \citep{esser_2021, yu_2021}.
Our goal is to characterize how tokenization choices influence learned representations and their relation to scientific inference.

\section{AstroPT backbone, tokenization strategies, and experimental setup}

We adopt AstroPT \citep{ref_smith2024} as our shared backbone:
AstroPT is a decoder-only transformer \citep{ref_vaswani2017,ref_radford2019} trained  autoregressively on sequences of astronomical data.
We choose an autoregressive, decoder-only architecture because it is the dominant paradigm in current foundation models and imposes minimal inductive bias on the representation, making it a neutral testbed for comparing tokenization strategies.
Our four tokenization strategies span a range of design choices: linear versus non-linear mappings, continuous versus discrete representations, and deterministic versus probabilistic decoding.
We go into further detail about each strategy below.

\textbf{The Affine tokenizer} defines the simplest possible mapping; a linear projection from image patches to the  embedding space, with a second linear projection for reconstruction. This makes Affine the minimal baseline in our suite: a single learned linear map in and out, with all representational work left to the backbone.

\textbf{The AIM tokenizer} replaces the linear projection with an MLP, following \citep{ref_elnouby2024} and the original AstroPT framework. The MLP head gives the tokeniser a non-linear, learnable mapping from patch pixels to the embedding space, in principle letting it capture structure within a patch that a single linear projection cannot.
Following \citep{ref_elnouby2024} and \citep{ref_smith2024}, we apply $z$-score normalization to each Affine and AIM tokenizer patch to stabilize training.

\textbf{The JetFormer tokenizer}, motivated by \citep{ref_tschannen2024}, produces continuous tokens through an invertible normalising flow. Inputs are uniformly dequantised, then a two-dimensional flow produces a latent image preserving all input information. Because the flow is invertible, the same module acts as both encoder and decoder and requires no separately pretrained tokeniser, instead being trained end-to-end to maximise the likelihood of the raw images. These continuous tokens are modelled with a Gaussian-mixture prediction head rather than a categorical distribution over a discrete codebook, placing JetFormer at the continuous end of our design axis.

\textbf{The VQ-VAE tokenizer} discretizes images into a learned codebook of visual tokens \citep{Oord_2017}.
An encoder network maps image patches to a continuous latent space, where each latent vector is quantized to its nearest neighbor in a learned codebook.
Following the literature, our VQ-VAE tokenizer is pre-trained on in-distribution galaxy imagery and then frozen during AstroPT training.

We must note here that our tokenization comparison is not entirely ablative, and that this is a genuine methodological tension.
However, we believe it is endemic to any realistic comparison of tokenizers: each strategy requires its own training procedure to be evaluated competitively, and forcing an entirely uniform regime would produce unrealistic and unfavorable setting for some methods.
We therefore instead follow the training methodologies described in the literature as faithfully as possible.

\textbf{Data selection and pre-processing.}
We use a publicly available galaxy imaging dataset\footnote{\url{https://hf.co/datasets/Smith42/galaxies}} derived from the DESI Legacy Surveys (DESI-LS) Data Release~8 \citep{ref_dey2019,ref_walmsley2023}, comprising $8.6$ million galaxies, from which we select a representative subset of $640\,000$ galaxies for training. DESI-LS was constructed to provide the targeting imagery for the DESI spectroscopic survey, and as such offers homogeneous, wide-field coverage well suited to large-scale representation learning.
For each galaxy we extract $256 \times 256$ pixel $g,r,z$--band postage stamps at a resolution of $0.262~\text{arcsec\,pixel}^{-1}$. The DR8 photometry is derived from model-based source fitting with the Tractor pipeline, which provides the magnitudes and colours we use as probe targets \citep{ref_dey2019}.

\textbf{Hyperparameters and pre-training routines.}
All experiments employ an $89$M-parameter AstroPT backbone\footnote{\url{https://github.com/Smith42/astroPT}} with $12$ transformer layers, model dimension $768$, and $12$ attention heads, with a context length of $1024$ tokens, causal self-attention, zero dropout, and bias-free linear layers.
Input images of size $256 \times 256$ are divided into $8 \times 8$ pixel patches and serialized in raster-scan order, yielding $1024$ tokens per image.
All tokenizers are optimized with AdamW \citep{ref_loshchilov2017} using a base learning rate of $10^{-4}$, weight decay of $10^{-4}$, $\beta_2 = 0.95$, batch size $8$, and gradient clipping at $0.5$, with a linear warm-up over $10\,000$ steps.

For JetFormer, we use $16$ flow steps and a GMM output head with $256$ components, following the progressive noise curriculum of \citet{ref_tschannen2024} in which RGB noise decays from $\sigma = 64$ to zero and latent noise from $0.3$ to zero over training. The noise curriculum is not incidental; without it the flow can encode information in imperceptible high-frequency dimensions and collapse to degenerate solutions.

The VQ-VAE tokenizer uses a codebook of $512$ embeddings of dimension $64$ with commitment loss $\lambda = 0.25$, an encoder-decoder hidden dimension of $128$ with $2$ residual layers \citep{ref_resnet}, trained for $50\,000$ steps with cross-entropy loss for next-token prediction on the AstroPT backbone.
Our pre-training code for all tokenizers is publicly available on the AstroPT Github repository.

\textbf{Physical knowledge and reconstruction performance tests.}
We train linear and MLP probes on embeddings extracted from the center layer of each AstroPT model to predict physical properties of 167\,000 held out galaxies.
We select a range of properties covering photometry, morphology, and spectroscopy.
Each property is evaluated with a $k$=10-fold cross validation.
Reconstruction quality is assessed via SSIM and PSNR on 5\,000 random test set galaxies. Because probes recover only information that is linearly or simply non-linearly accessible, they measure how a representation organises its physical content rather than whether that content is present at all---a distinction we return to in interpreting JetFormer.


\section{Does tokenization affect latent physical knowledge?}

We evaluate whether tokenization affects (a) how much physical information is retained in the learned representations, (b) how that information is organized (linearly vs. nonlinearly), and (c) whether reconstruction quality predicts either of these.
Table~\ref{fig:results} summarizes our main results across linear and MLP probes for physical property prediction, as well as SSIM and PSNR reconstruction metrics. The probed properties include DESI photometric magnitudes ($M_g, M_r$) and colors ($g-r$, $r-z$), photometric and spectroscopic redshifts (photo-$z$, spec-$z$), specific star formation rate (sSFR), stellar mass $M_*$, and morphological indicators (smoothness, disk fraction, artifact fraction, edge-on fraction, and spiral winding tightness).

\begin{table}[h]
    \centering
    \caption{Model evaluation results. Linear and MLP probe $R^2$ for physical properties. SSIM and $R^2$ values are multiplied by a factor of 100.}
    \label{fig:results}
    \includegraphics[width=\columnwidth]{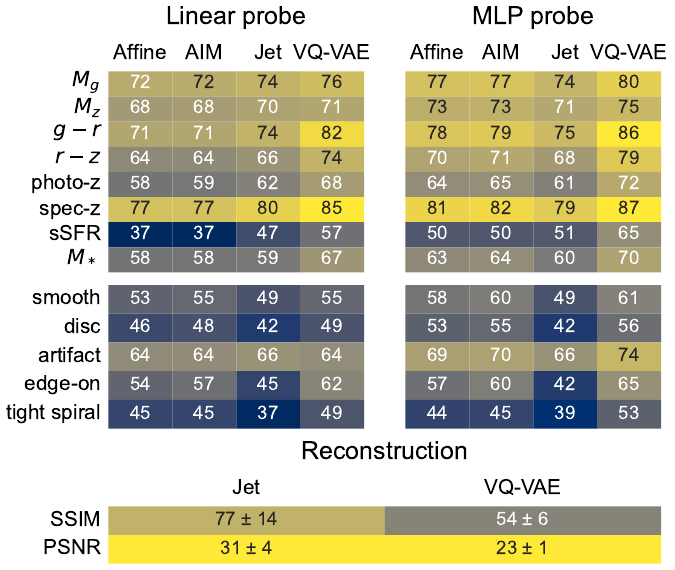}
\end{table}

\subsection{Regression on galaxy physical parameters.}
VQ-VAE achieves the strongest linear and MLP probe performance for most physical properties.
Notably, VQ-VAE's linear probe $R^2$ surpasses the MLP probe $R^2$ for most properties, which suggests that the discrete bottleneck organises physical information in a more linearly accessible manner compared to the other tokenizers.

JetFormer consistently outperforms AIM and Affine on linear probes for color and spectrally related quantities (magnitudes, colors, redshifts, star formation rate, stellar mass), while AIM shows stronger performance on morphology-driven parameters (smoothness, disk fraction, edge-on fraction).
Interestingly, the linear and MLP probe performances for JetFormer are within the error spread from each other, suggesting that the JetFormer's normalizing flow produces embeddings that are somewhat linearly accessible, with little-to-no additional information accessible via a non-linear method. Affine and AIM perform near-identically across nearly all properties.
This indicates that the MLP in the AIM tokenizer head adds negligible benefit over a simple linear projection when the transformer backbone is sufficiently expressive.

Across all tokenizers and probe types, $g-r$ color is predicted more accurately than $r-z$.
This is initially surprising given that $g-r$ is more susceptible to dust obscuration, and the associated age-dust degeneracy (an old galaxy with no dust can have the same $g-r$ color as a young, star-forming galaxy that is heavily obscured).
However, $g-r$ spans a larger dynamical range in our redshift regime and varies more strongly over the training set \citep{ref_raichoor2023}, making it an easier regression target.
Physically, $g-r$ traces specific star formation rate (sSFR)via emission from young stars, whereas $r-z$ is a better predictor of older stellar populations that form the bulk of the stellar population, making it a better predictor of the total stellar mass, $M_*$.
The greater dust sensitivity of $g-r$ explains why sSFR, is harder to predict than $M_*$, and why the MLP probe shows a larger improvement over the linear probe for sSFR; non-linear degeneracies between dust, age, and color are better captured by the more expressive model.

The observed differences in performance likely reflect architectural design.
We hypothesize that JetFormer's multiple non-linear transformations emphasize global and spectrally informative features, but may smooth fine-scale morphological information. Patch-based methods (AIM, Affine), on the other hand, preserve localized spatial information, aiding encoding of morphological features. Finally, VQ-VAE further enforces a discrete information bottleneck through its finite codebook, compressing image patches into a small set of representative tokens. This hard tokenization discards nuisance variation while encouraging semantic clustering aligned with underlying physical properties, resulting in embeddings that are particularly well suited for regression tasks.

\textbf{To investigate how each tokenizer allocates its representational capacity}, we perform a second run of $k=10$-fold linear probes to predict our embedding spaces given groups of physical properties.
We find that JetFormer dedicates larger shares of its representational capacity to directly observable image properties, with the `apparent photometry' and `structural parameter' group probes reaching $R^2 = 0.33\pm0.02$ and $R^2 = 0.29\pm0.02$, compared to VQ-VAE's $R^2 = 0.19\pm0.01$ and $R^2 = 0.16\pm0.01$.
Conversely, VQ-VAE dedicates more capacity `higher level' physical properties that require some abstraction beyond pixelwise analysis such as redshift ($R^2 = 0.058\pm0.003$ vs $R^2 = 0.033\pm0.006$), and absolute photometry ($R^2 = 0.090\pm0.004$ vs $R^2 = 0.05 \pm 0.01$).
This suggests that the VQ-VAE tokenizer enables higher level reasoning which allows our AstroPT backbone to dedicate more space to complex physical properties.

\subsection{Image reconstruction.}

JetFormer achieves the highest reconstruction fidelity, preserving fine spatial details and low surface brightness components more effectively than other approaches (Fig.~\ref{fig:tokenizer_reconstruction_comparison}).
On 5\,000 held-out images JetFormer achieves a mean PSNR of $31.11\,\mathrm{dB}$ ($\sigma = 3.54$) and mean SSIM of $0.762$ ($\sigma = 0.138$), confirming that it preserves structural morphologies (spiral arms, galactic cores, and diffuse background emission).
For comparison, VQ-VAE attains a substantially lower mean PSNR of $23.57\,\mathrm{dB}$ ($\sigma = 0.93$) and SSIM of $0.544$ ($\sigma = 0.054$), quantitatively supporting the reconstruction patterns observed in Figure~\ref{fig:tokenizer_reconstruction_comparison}.
VQ-VAE produces visually plausible reconstructions but exhibits diffuse or `cloudy' artifacts surrounding bright cores and spurious localized features (small red structures absent in the input data).
It also underperforms on low-surface-brightness backgrounds, indicating limited recovery of faint, extended emission.
In contrast, JetFormer generates consistently sharper reconstructions with improved contrast and structural coherence.

For AIM and Affine, patch-level $z$-score normalization produces a block-wise appearance in reconstruction.
This normalization enhances training stability and contributes to strong probe performance, but limits recovery of fine-grained intensity variations in the reconstructed images.
Reconstructed images from these methods may visually differ from those of other methods despite sharing the same underlying target.
Taking our probe and reconstruction results together, these findings highlight a trade-off between information preservation and abstraction: while soft, dense, information-complete representations favor high-fidelity reconstruction, hard and patch-based tokenization strategies promote more directly interpretable and easily accessible representations for physical inference.

\textbf{Probing the reconstructed images.}
JetFormer attains the highest reconstruction fidelity yet yields the least linearly accessible embedding among continuous tokenizers, while VQ--VAE shows the opposite.
To test whether this reflects genuine information loss or merely inaccessible information, we probe the reconstructed images themselves rather than the AstroPT embeddings.
For each of the $180\,000$ test and validation galaxies, we extract ImageNet-pretrained ResNet-50 features \citep{he2016deep} from the JetFormer reconstruction, the VQ--VAE reconstruction, and the original image, then repeat the linear and MLP probes on all three. 
Holding the feature
extractor, the input resolution, and the probing protocol static across tests isolates the
\emph{information content of the reconstructed pixels}.
Our targets are the eight photometric and five morphological properties from Tab.~\ref{fig:results}.

\begin{table}[h]
    \centering
    \caption{Linear (\textit{left}) and MLP (\textit{right}) probing of
  ImageNet-pretrained ResNet-50 features extracted from (i) the original
  galaxy images, (ii) JetFormer reconstructions, and (iii) VQ--VAE
  reconstructions. Cells report $R^2 \times 100$ for the prediction of
  eight photometric properties (top) and five morphological properties
  (bottom). JetFormer reconstructions match the Original baseline to within
  a few $R^2$ points on nearly every target, while VQ--VAE reconstructions
  show a systematic information loss that is largest on absolute photometry
  and morphology.}
    \label{fig:results2}
    \includegraphics[width=\columnwidth]{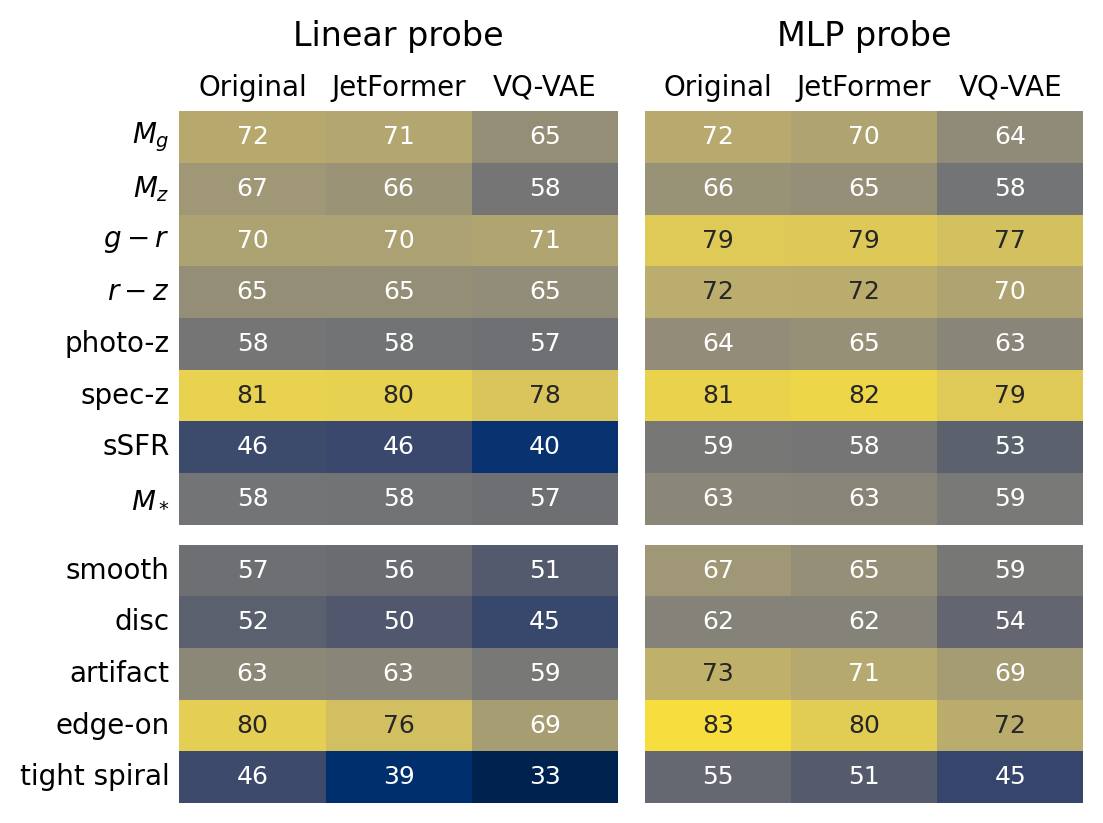}
\end{table}

We find that our JetFormer reconstructions retain nearly all of the information present in the original imagery, matching the original-image baseline to within error on every photometric target and lagging by at most four points on morphology (Tab.~\ref{fig:results2}).
In contrast, the VQ--VAE reconstructions lose fine information systematically (such as in morphology, sSFR, and absolute photometry) while maintaining colour indices, which are insensitive to a multiplicative
offset of the surface-brightness profile. 
This confirms that VQ--VAE's codebook discards low-surface-brightness structure, fine morphology, and absolute photometric scale.
To summarize, we find that VQ--VAE places its physical content directly in an accessible embedding geometry but irrevocably removes pixel detail, while JetFormer carries the full input through to its pixels but scatters it in the embedding, where neither MLP nor linear probes are powerful enough to recover individual physical properties.

\section{Trade-offs, limitations, and the case for physics-grounded benchmarks}

Our comparison of four tokenization strategies within a shared AstroPT backbone shows that tokenization is not a neutral preprocessing step, but shapes what information is learned and how it is organized within the representation.
We find a decoupling between reconstruction fidelity and the physical information accessible in the representation: JetFormer achieves the best reconstruction, while VQ-VAE yields the most informative embeddings for physical property prediction.
Probing the reconstructed images directly establishes the mechanism behind this decoupling. VQ-VAE's discrete bottleneck discards pixel-level detail yet places physical content directly in the geometry of the embedding, whereas JetFormer's invertible flow preserves the input information through to the reconstructed pixels but scatters it across the latent space in a form that simple probes cannot surface. Reconstruction fidelity and probe-accessible representation quality are therefore not merely uncorrelated but governed by distinct underlying mechanisms.
Patch-based methods (AIM, Affine) occupy a middle ground, performing well on morphology-driven features and showing that the transformer backbone largely drives representation learning in this regime.

These trade-offs suggest that tokenization should be chosen based on performance on the eventual downstream task rather than reconstruction metrics alone.
A model intended for photometric or spectroscopic inference may benefit from VQ-VAE's compressed representations, while one targeting image generation or reconstruction should prefer JetFormer's approach, which we have shown to be an information-preserving channel.
For compute-constrained settings, the Affine tokenizer achieves performance comparable to AIM at lower complexity (Appendix~\ref{app:carbon}).

We note several limitations.
Our embedding-level evaluation relies on linear and MLP probes, which may understate the information present in the representation; while probing the reconstructions mitigates this concern for the two extreme cases---confirming that JetFormer's information is genuinely present but not readily accessible---a fully probe-independent comparison across all four tokenizers remains open.
Differences in tokenizer training procedures mean the comparison is not purely ablative, and VQ-VAE's performance may partly reflect its pre-trained encoder. JetFormer sits at the opposite end of this confound, having been designed specifically to remove the pretrained-encoder dependency that complicates the VQ-VAE comparison.
We also do not explore scaling with dataset size or model capacity, and patch normalization complicates direct reconstruction comparisons.

To summarize, astronomy provides a powerful setting for tokenizer analysis: known physics and independently measured quantities offer objective benchmarks that avoid the circularity of evaluating representations solely on human-annotated tasks or toy datasets.
We expect that similar evaluation frameworks would be valuable in other domains where physical ground truth is likewise available; future work will lay the groundwork for this by extending this study's framework towards realizing a comprehensive `tokenization bench' built on verifiable astronomical ground-truth data.

\section*{Acknowledgments}
We would like to thank Stella Biderman and EleutherAI for their generous donation of compute which made this project possible.

We would like to thank the AstroAI group at the Center for Astrophysics \textbar{} Harvard \& Smithsonian for illuminating discussion and comments that improved this paper.

\bibliographystyle{icml2026}
\bibliography{main}

\begin{thebibliography}{31}
\providecommand{\natexlab}[1]{#1}
\providecommand{\url}[1]{\texttt{#1}}
\expandafter\ifx\csname urlstyle\endcsname\relax
  \providecommand{\doi}[1]{doi: #1}\else
  \providecommand{\doi}{doi: \begingroup \urlstyle{rm}\Url}\fi

\bibitem[{Amiaux} et~al.(2012){Amiaux}, {Scaramella}, {Mellier}, {Altieri}, {Burigana}, {Da Silva}, {Gomez}, {Hoar}, {Laureijs}, {Maiorano}, {Magalh{\~a}es Oliveira}, {Renk}, {Saavedra Criado}, {Tereno}, {Augu{\`e}res}, {Brinchmann}, {Cropper}, {Duvet}, {Ealet}, {Franzetti}, {Garilli}, {Gondoin}, {Guzzo}, {Hoekstra}, {Holmes}, {Jahnke}, {Kitching}, {Meneghetti}, {Percival}, and {Warren}]{ref_euclid}
{Amiaux}, J., {Scaramella}, R., {Mellier}, Y., {Altieri}, B., {Burigana}, C., {Da Silva}, A., {Gomez}, P., {Hoar}, J., {Laureijs}, R., {Maiorano}, E., {Magalh{\~a}es Oliveira}, D., {Renk}, F., {Saavedra Criado}, G., {Tereno}, I., {Augu{\`e}res}, J.~L., {Brinchmann}, J., {Cropper}, M., {Duvet}, L., {Ealet}, A., {Franzetti}, P., {Garilli}, B., {Gondoin}, P., {Guzzo}, L., {Hoekstra}, H., {Holmes}, R., {Jahnke}, K., {Kitching}, T., {Meneghetti}, M., {Percival}, W., and {Warren}, S.
\newblock {Euclid mission: building of a reference survey}.
\newblock In {Clampin}, M.~C., {Fazio}, G.~G., {MacEwen}, H.~A., and {Oschmann}, Jacobus~M., J. (eds.), \emph{Space Telescopes and Instrumentation 2012: Optical, Infrared, and Millimeter Wave}, volume 8442 of \emph{Society of Photo-Optical Instrumentation Engineers (SPIE) Conference Series}, pp.\  84420Z, 2012.
\newblock \doi{10.1117/12.926513}.

\bibitem[{Black Forest Labs}(2025)]{ref_bfl_representation}
{Black Forest Labs}.
\newblock {FLUX.2}: Analyzing and enhancing the latent space of {FLUX} -- representation comparison, 2025.
\newblock URL \url{https://bfl.ai/research/representation-comparison}.

\bibitem[{Dey} et~al.(2019){Dey}, {Schlegel}, {Lang}, {Blum}, {Burleigh}, {Fan}, {Findlay}, {Finkbeiner}, {Herrera}, {Juneau}, {Landriau}, {Levi}, {McGreer}, {Meisner}, {Myers}, {Moustakas}, {Nugent}, {Patej}, {Schlafly}, {Walker}, {Valdes}, {Weaver}, {Y{\`e}che}, {Zou}, {Zhou}, {Abareshi}, {Abbott}, {Abolfathi}, {Aguilera}, {Alam}, {Allen}, {Alvarez}, {Annis}, {Ansarinejad}, {Aubert}, {Beechert}, {Bell}, {BenZvi}, {Beutler}, {Bielby}, {Bolton}, {Brice{\~n}o}, {Buckley-Geer}, {Butler}, {Calamida}, {Carlberg}, {Carter}, {Casas}, {Castander}, {Choi}, {Comparat}, {Cukanovaite}, {Delubac}, {DeVries}, {Dey}, {Dhungana}, {Dickinson}, {Ding}, {Donaldson}, {Duan}, {Duckworth}, {Eftekharzadeh}, {Eisenstein}, {Etourneau}, {Fagrelius}, {Farihi}, {Fitzpatrick}, {Font-Ribera}, {Fulmer}, {G{\"a}nsicke}, {Gaztanaga}, {George}, {Gerdes}, {Gontcho}, {Gorgoni}, {Green}, {Guy}, {Harmer}, {Hernandez}, {Honscheid}, {Huang}, {James}, {Jannuzi}, {Jiang}, {Joyce}, {Karcher}, {Karkar}, {Kehoe}, {Kneib}, {Kueter-Young}, {Lan},
  {Lauer}, {Le Guillou}, {Le Van Suu}, {Lee}, {Lesser}, {Perreault Levasseur}, {Li}, {Mann}, {Marshall}, {Mart{\'\i}nez-V{\'a}zquez}, {Martini}, {du Mas des Bourboux}, {McManus}, {Meier}, {M{\'e}nard}, {Metcalfe}, {Mu{\~n}oz-Guti{\'e}rrez}, {Najita}, {Napier}, {Narayan}, {Newman}, {Nie}, {Nord}, {Norman}, {Olsen}, {Paat}, {Palanque-Delabrouille}, {Peng}, {Poppett}, {Poremba}, {Prakash}, {Rabinowitz}, {Raichoor}, {Rezaie}, {Robertson}, {Roe}, {Ross}, {Ross}, {Rudnick}, {Safonova}, {Saha}, {S{\'a}nchez}, {Savary}, {Schweiker}, {Scott}, {Seo}, {Shan}, {Silva}, {Slepian}, {Soto}, {Sprayberry}, {Staten}, {Stillman}, {Stupak}, {Summers}, {Sien Tie}, {Tirado}, {Vargas-Maga{\~n}a}, {Vivas}, {Wechsler}, {Williams}, {Yang}, {Yang}, {Yapici}, {Zaritsky}, {Zenteno}, {Zhang}, {Zhang}, {Zhou}, and {Zhou}]{ref_desi}
{Dey}, A., {Schlegel}, D.~J., {Lang}, D., {Blum}, R., {Burleigh}, K., {Fan}, X., {Findlay}, J.~R., {Finkbeiner}, D., {Herrera}, D., {Juneau}, S., {Landriau}, M., {Levi}, M., {McGreer}, I., {Meisner}, A., {Myers}, A.~D., {Moustakas}, J., {Nugent}, P., {Patej}, A., {Schlafly}, E.~F., {Walker}, A.~R., {Valdes}, F., {Weaver}, B.~A., {Y{\`e}che}, C., {Zou}, H., {Zhou}, X., {Abareshi}, B., {Abbott}, T., {Abolfathi}, B., {Aguilera}, C., {Alam}, S., {Allen}, L., {Alvarez}, A., {Annis}, J., {Ansarinejad}, B., {Aubert}, M., {Beechert}, J., {Bell}, E.~F., {BenZvi}, S.~Y., {Beutler}, F., {Bielby}, R.~M., {Bolton}, A.~S., {Brice{\~n}o}, C., {Buckley-Geer}, E.~J., {Butler}, K., {Calamida}, A., {Carlberg}, R.~G., {Carter}, P., {Casas}, R., {Castander}, F.~J., {Choi}, Y., {Comparat}, J., {Cukanovaite}, E., {Delubac}, T., {DeVries}, K., {Dey}, S., {Dhungana}, G., {Dickinson}, M., {Ding}, Z., {Donaldson}, J.~B., {Duan}, Y., {Duckworth}, C.~J., {Eftekharzadeh}, S., {Eisenstein}, D.~J., {Etourneau}, T., {Fagrelius}, P.~A.,
  {Farihi}, J., {Fitzpatrick}, M., {Font-Ribera}, A., {Fulmer}, L., {G{\"a}nsicke}, B.~T., {Gaztanaga}, E., {George}, K., {Gerdes}, D.~W., {Gontcho}, S. G.~A., {Gorgoni}, C., {Green}, G., {Guy}, J., {Harmer}, D., {Hernandez}, M., {Honscheid}, K., {Huang}, L.~W., {James}, D.~J., {Jannuzi}, B.~T., {Jiang}, L., {Joyce}, R., {Karcher}, A., {Karkar}, S., {Kehoe}, R., {Kneib}, J., {Kueter-Young}, A., {Lan}, T., {Lauer}, T.~R., {Le Guillou}, L., {Le Van Suu}, A., {Lee}, J.~H., {Lesser}, M., {Perreault Levasseur}, L., {Li}, T.~S., {Mann}, J.~L., {Marshall}, R., {Mart{\'\i}nez-V{\'a}zquez}, C., {Martini}, P., {du Mas des Bourboux}, H., {McManus}, S., {Meier}, T.~G., {M{\'e}nard}, B., {Metcalfe}, N., {Mu{\~n}oz-Guti{\'e}rrez}, A., {Najita}, J., {Napier}, K., {Narayan}, G., {Newman}, J.~A., {Nie}, J., {Nord}, B., {Norman}, D.~J., {Olsen}, K. A.~G., {Paat}, A., {Palanque-Delabrouille}, N., {Peng}, X., {Poppett}, C.~L., {Poremba}, M.~R., {Prakash}, A., {Rabinowitz}, D., {Raichoor}, A., {Rezaie}, M., {Robertson}, A.,
  {Roe}, N.~A., {Ross}, A.~J., {Ross}, N.~P., {Rudnick}, G., {Safonova}, S., {Saha}, A., {S{\'a}nchez}, F.~J., {Savary}, E., {Schweiker}, H., {Scott}, A., {Seo}, H., {Shan}, H., {Silva}, D.~R., {Slepian}, Z., {Soto}, C., {Sprayberry}, D., {Staten}, R., {Stillman}, C.~M., {Stupak}, R.~J., {Summers}, D.~L., {Sien Tie}, S., {Tirado}, H., {Vargas-Maga{\~n}a}, M., {Vivas}, A.~K., {Wechsler}, R.~H., {Williams}, D., {Yang}, J., {Yang}, Q., {Yapici}, T., {Zaritsky}, D., {Zenteno}, A., {Zhang}, K., {Zhang}, T., {Zhou}, R., and {Zhou}, Z.
\newblock Overview of the desi legacy imaging surveys.
\newblock \emph{The Astronomical Journal}, 157\penalty0 (5):\penalty0 168, 2019.
\newblock \doi{10.3847/1538-3881/ab089d}.

\bibitem[Dey et~al.(2019)Dey, Schlegel, Lang, Blum, Burleigh, Fan, Findlay, Finkbeiner, Herrera, Juneau, et~al.]{ref_dey2019}
Dey, A., Schlegel, D.~J., Lang, D., Blum, R., Burleigh, K., Fan, X., Findlay, J.~R., Finkbeiner, D., Herrera, D., Juneau, S., et~al.
\newblock {Overview of the DESI Legacy Imaging Surveys}.
\newblock \emph{Astronomical Journal}, 157\penalty0 (5):\penalty0 168, 2019.
\newblock ISSN 1538-3881.
\newblock \doi{10.3847/1538-3881/ab089d}.

\bibitem[El-Nouby et~al.(2024)El-Nouby, Klein, Zhai, Bautista, Toshev, Shankar, Susskind, and Joulin]{ref_elnouby2024}
El-Nouby, A., Klein, M., Zhai, S., Bautista, M.~A., Toshev, A., Shankar, V., Susskind, J.~M., and Joulin, A.
\newblock {Scalable Pre-training of Large Autoregressive Image Models}.
\newblock \emph{ArXiv e-prints}, 2024.
\newblock \doi{10.48550/arXiv.2401.08541}.

\bibitem[Esser et~al.(2021)Esser, Rombach, and Ommer]{esser_2021}
Esser, P., Rombach, R., and Ommer, B.
\newblock Taming transformers for high-resolution image synthesis.
\newblock In \emph{Proceedings of the IEEE/CVF conference on computer vision and pattern recognition}, pp.\  12873--12883, 2021.

\bibitem[{Euclid Collaboration} et~al.(2025){Euclid Collaboration}, Siudek, Huertas-Company, Smith, Martinez-Solaeche, Lanusse, Ho, Angeloudi, Cunha, S{\ifmmode\acute{a}\else\'{a}\fi}nchez, Dunn, Fu, Iglesias-Navarro, Junais, Knapen, Laloux, Mezcua, Roster, Stevens, Vega-Ferrero, Aghanim, Altieri, Amara, Andreon, Auricchio, Aussel, Baccigalupi, Baldi, Bardelli, Battaglia, Biviano, Bonchi, Branchini, Brescia, Brinchmann, Camera, Ca{\ifmmode\tilde{n}\else\~{n}\fi}as-Herrera, Capobianco, Carbone, Carretero, Casas, Castander, Castellano, Castignani, Cavuoti, Chambers, Cimatti, Colodro-Conde, Congedo, Conselice, Conversi, Copin, Courbin, Courtois, Cropper, Da~Silva, Degaudenzi, De~Lucia, Di~Giorgio, Dinis, Dolding, Dole, Dubath, Duncan, Dupac, Dusini, Escoffier, Farina, Farinelli, Faustini, Ferriol, Finelli, Fotopoulou, Frailis, Franceschi, Galeotta, George, Gillis, Giocoli, Gracia-Carpio, Granett, Grazian, Grupp, Gwyn, Haugan, Holmes, Hook, Hormuth, Hornstrup, Jahnke, Jhabvala,
  Keih{\ifmmode\ddot{a}\else\"{a}\fi}nen, Kermiche, Kiessling, Kubik, K{\ifmmode\ddot{u}\else\"{u}\fi}mmel, Kunz, Kurki-Suonio, Boulc'h, Brun, Mignant, Ligori, Lilje, Lindholm, Lloro, Mainetti, Maino, Maiorano, Mansutti, Marcin, Marggraf, Martinelli, Martinet, Marulli, Massey, Maurogordato, McCracken, Medinaceli, Mei, Melchior, Mellier, Meneghetti, Merlin, Meylan, Mora, Moresco, Moscardini, Nakajima, Neissner, Niemi, Nightingale, Padilla, Paltani, Pasian, Pedersen, Percival, Pettorino, Pires, Polenta, Poncet, Popa, Pozzetti, Raison, Renzi, Rhodes, Riccio, Romelli, Roncarelli, Saglia, Sakr, S{\ifmmode\acute{a}\else\'{a}\fi}nchez, Sapone, Sartoris, Schewtschenko, Schneider, Schrabback, Scodeggio, Secroun, Seidel, Seiffert, Serrano, Simon, Sirignano, Sirri, Stanco, Steinwagner, Tallada-Cresp{\ifmmode\acute{\imath}\else\'{\i}\fi}, Taylor, Tereno, Toft, Toledo-Moreo, Torradeflot, Tutusaus, Valenziano, Valiviita, Vassallo, Kleijn, Veropalumbo, Wang, Weller, Zacchei, Zamorani, Zerbi, Zinchenko, Zucca, Allevato,
  Ballardini, Bolzonella, Bozzo, Burigana, Cabanac, Cappi, Di~Ferdinando, Vigo, Gabarra, Mart{\ifmmode\acute{\imath}\else\'{\i}\fi}n-Fleitas, Matthew, Mauri, Metcalf, Pezzotta, P{\ifmmode\ddot{o}\else\"{o}\fi}ntinen, Porciani, Risso, Scottez, Sereno, Tenti, Viel, Wiesmann, Akrami, Andika, Anselmi, Archidiacono, Atrio-Barandela, Benoist, Benson, Bertacca, Bethermin, Bisigello, Blanchard, Blot, Brown, Bruton, Calabro, Quevedo, Caro, Carvalho, Castro, Charles, Cogato, Cooray, Cucciati, Davini, De~Paolis, Desprez, D{\ifmmode\acute{\imath}\else\'{\i}\fi}az-S{\ifmmode\acute{a}\else\'{a}\fi}nchez, Diaz, Di~Domizio, Diego, Duc, Enia, Fang, Ferrari, Ferreira, Finoguenov, Fontana, Franco, Ganga, Garc{\ifmmode\acute{\imath}\else\'{\i}\fi}a-Bellido, Gasparetto, Gautard, Gaztanaga, Giacomini, Gianotti, Gozaliasl, Guidi, Gutierrez, Hall, Hartley, Hemmati, Hern{\ifmmode\acute{a}\else\'{a}\fi}ndez-Monteagudo, Hildebrandt, Hjorth, Kajava, Kang, Kansal, Karagiannis, Kiiveri, Kirkpatrick, Kruk, Graet, Legrand, Lembo, Lepori,
  Leroy, Lesci, Lesgourgues, Leuzzi, Liaudat, Loureiro, Macias-Perez, Maggio, Magliocchetti, Magnier, Mannucci, Maoli, Martins, Maurin, Miluzio, Monaco, Moretti, Morgante, Murray, Naidoo, Navarro-Alsina, Nesseris, Passalacqua, Paterson, Patrizii, Pisani, Potter, Quai, Radovich, Sacquegna, Sahl{\ifmmode\acute{e}\else\'{e}\fi}n, Sanders, Sarpa, Schneider, Sciotti, Scognamiglio, Sellentin, Smith, Tanidis, Testera, Teyssier, Tosi, Troja, Tucci, Valieri, Venhola, Vergani, Verza, Vielzeuf, Walton, and Sorce]{ref_euclid2025}
{Euclid Collaboration}, Siudek, M., Huertas-Company, M., Smith, M., Martinez-Solaeche, G., Lanusse, F., Ho, S., Angeloudi, E., Cunha, P. A.~C., S{\ifmmode\acute{a}\else\'{a}\fi}nchez, H.~D., Dunn, M., Fu, Y., Iglesias-Navarro, P., Junais, J., Knapen, J.~H., Laloux, B., Mezcua, M., Roster, W., Stevens, G., Vega-Ferrero, J., Aghanim, N., Altieri, B., Amara, A., Andreon, S., Auricchio, N., Aussel, H., Baccigalupi, C., Baldi, M., Bardelli, S., Battaglia, P., Biviano, A., Bonchi, A., Branchini, E., Brescia, M., Brinchmann, J., Camera, S., Ca{\ifmmode\tilde{n}\else\~{n}\fi}as-Herrera, G., Capobianco, V., Carbone, C., Carretero, J., Casas, S., Castander, F.~J., Castellano, M., Castignani, G., Cavuoti, S., Chambers, K.~C., Cimatti, A., Colodro-Conde, C., Congedo, G., Conselice, C.~J., Conversi, L., Copin, Y., Courbin, F., Courtois, H.~M., Cropper, M., Da~Silva, A., Degaudenzi, H., De~Lucia, G., Di~Giorgio, A.~M., Dinis, J., Dolding, C., Dole, H., Dubath, F., Duncan, C. A.~J., Dupac, X., Dusini, S., Escoffier, S.,
  Farina, M., Farinelli, R., Faustini, F., Ferriol, S., Finelli, F., Fotopoulou, S., Frailis, M., Franceschi, E., Galeotta, S., George, K., Gillis, B., Giocoli, C., Gracia-Carpio, J., Granett, B.~R., Grazian, A., Grupp, F., Gwyn, S., Haugan, S. V.~H., Holmes, W., Hook, I.~M., Hormuth, F., Hornstrup, A., Jahnke, K., Jhabvala, M., Keih{\ifmmode\ddot{a}\else\"{a}\fi}nen, E., Kermiche, S., Kiessling, A., Kubik, B., K{\ifmmode\ddot{u}\else\"{u}\fi}mmel, M., Kunz, M., Kurki-Suonio, H., Boulc'h, Q.~L., Brun, A. M. C.~L., Mignant, D.~L., Ligori, S., Lilje, P.~B., Lindholm, V., Lloro, I., Mainetti, G., Maino, D., Maiorano, E., Mansutti, O., Marcin, S., Marggraf, O., Martinelli, M., Martinet, N., Marulli, F., Massey, R., Maurogordato, S., McCracken, H.~J., Medinaceli, E., Mei, S., Melchior, M., Mellier, Y., Meneghetti, M., Merlin, E., Meylan, G., Mora, A., Moresco, M., Moscardini, L., Nakajima, R., Neissner, C., Niemi, S.-M., Nightingale, J.~W., Padilla, C., Paltani, S., Pasian, F., Pedersen, K., Percival, W.~J.,
  Pettorino, V., Pires, S., Polenta, G., Poncet, M., Popa, L.~A., Pozzetti, L., Raison, F., Renzi, A., Rhodes, J., Riccio, G., Romelli, E., Roncarelli, M., Saglia, R., Sakr, Z., S{\ifmmode\acute{a}\else\'{a}\fi}nchez, A.~G., Sapone, D., Sartoris, B., Schewtschenko, J.~A., Schneider, P., Schrabback, T., Scodeggio, M., Secroun, A., Seidel, G., Seiffert, M., Serrano, S., Simon, P., Sirignano, C., Sirri, G., Stanco, L., Steinwagner, J., Tallada-Cresp{\ifmmode\acute{\imath}\else\'{\i}\fi}, P., Taylor, A.~N., Tereno, I., Toft, S., Toledo-Moreo, R., Torradeflot, F., Tutusaus, I., Valenziano, L., Valiviita, J., Vassallo, T., Kleijn, G.~V., Veropalumbo, A., Wang, Y., Weller, J., Zacchei, A., Zamorani, G., Zerbi, F.~M., Zinchenko, I.~A., Zucca, E., Allevato, V., Ballardini, M., Bolzonella, M., Bozzo, E., Burigana, C., Cabanac, R., Cappi, A., Di~Ferdinando, D., Vigo, J. A.~E., Gabarra, L., Mart{\ifmmode\acute{\imath}\else\'{\i}\fi}n-Fleitas, J., Matthew, S., Mauri, N., Metcalf, R.~B., Pezzotta, A.,
  P{\ifmmode\ddot{o}\else\"{o}\fi}ntinen, M., Porciani, C., Risso, I., Scottez, V., Sereno, M., Tenti, M., Viel, M., Wiesmann, M., Akrami, Y., Andika, I.~T., Anselmi, S., Archidiacono, M., Atrio-Barandela, F., Benoist, C., Benson, K., Bertacca, D., Bethermin, M., Bisigello, L., Blanchard, A., Blot, L., Brown, M.~L., Bruton, S., Calabro, A., Quevedo, B.~C., Caro, F., Carvalho, C.~S., Castro, T., Charles, Y., Cogato, F., Cooray, A.~R., Cucciati, O., Davini, S., De~Paolis, F., Desprez, G., D{\ifmmode\acute{\imath}\else\'{\i}\fi}az-S{\ifmmode\acute{a}\else\'{a}\fi}nchez, A., Diaz, J.~J., Di~Domizio, S., Diego, J.~M., Duc, P.-A., Enia, A., Fang, Y., Ferrari, A.~G., Ferreira, P.~G., Finoguenov, A., Fontana, A., Franco, A., Ganga, K., Garc{\ifmmode\acute{\imath}\else\'{\i}\fi}a-Bellido, J., Gasparetto, T., Gautard, V., Gaztanaga, E., Giacomini, F., Gianotti, F., Gozaliasl, G., Guidi, M., Gutierrez, C.~M., Hall, A., Hartley, W.~G., Hemmati, S., Hern{\ifmmode\acute{a}\else\'{a}\fi}ndez-Monteagudo, C., Hildebrandt, H.,
  Hjorth, J., Kajava, J. J.~E., Kang, Y., Kansal, V., Karagiannis, D., Kiiveri, K., Kirkpatrick, C.~C., Kruk, S., Graet, J.~L., Legrand, L., Lembo, M., Lepori, F., Leroy, G., Lesci, G.~F., Lesgourgues, J., Leuzzi, L., Liaudat, T.~I., Loureiro, A., Macias-Perez, J., Maggio, G., Magliocchetti, M., Magnier, E.~A., Mannucci, F., Maoli, R., Martins, C. J. A.~P., Maurin, L., Miluzio, M., Monaco, P., Moretti, C., Morgante, G., Murray, C., Naidoo, K., Navarro-Alsina, A., Nesseris, S., Passalacqua, F., Paterson, K., Patrizii, L., Pisani, A., Potter, D., Quai, S., Radovich, M., Sacquegna, S., Sahl{\ifmmode\acute{e}\else\'{e}\fi}n, M., Sanders, D.~B., Sarpa, E., Schneider, A., Sciotti, D., Scognamiglio, D., Sellentin, E., Smith, L.~C., Tanidis, K., Testera, G., Teyssier, R., Tosi, S., Troja, A., Tucci, M., Valieri, C., Venhola, A., Vergani, D., Verza, G., Vielzeuf, P., Walton, N.~A., and Sorce, J.~G.
\newblock {Euclid Quick Data Release (Q1) Exploring galaxy properties with a multi-modal foundation model}.
\newblock \emph{ArXiv e-prints}, 2025.
\newblock \doi{10.48550/arXiv.2503.15312}.

\bibitem[He et~al.(2015)He, Zhang, Ren, and Sun]{ref_resnet}
He, K., Zhang, X., Ren, S., and Sun, J.
\newblock {Deep Residual Learning for Image Recognition}.
\newblock \emph{arXiv}, 2015.
\newblock \doi{10.48550/arXiv.1512.03385}.

\bibitem[He et~al.(2016)He, Zhang, Ren, and Sun]{he2016deep}
He, K., Zhang, X., Ren, S., and Sun, J.
\newblock Deep residual learning for image recognition.
\newblock In \emph{Proceedings of the IEEE Conference on Computer Vision and Pattern Recognition (CVPR)}, pp.\  770--778, 2016.

\bibitem[Hemmati et~al.(2026)Hemmati, Krick, Stern, Desai, Faisst, Martín-García, Gorjian, Haghjoo, Nikakhtar, Raen, Sanjaripour, Sipőcz, and Shupe]{Hemmati2026}
Hemmati, S., Krick, J., Stern, D., Desai, V., Faisst, A., Martín-García, L., Gorjian, V., Haghjoo, A., Nikakhtar, F., Raen, T., Sanjaripour, S., Sipőcz, B.~M., and Shupe, D.
\newblock Reducing the dimensions of active galactic nuclei light-curve manifolds.
\newblock \emph{The Astrophysical Journal}, 998\penalty0 (1):\penalty0 130, feb 2026.
\newblock \doi{10.3847/1538-4357/ae38b8}.
\newblock URL \url{https://doi.org/10.3847/1538-4357/ae38b8}.

\bibitem[{Ivezi{\'c}} et~al.(2019){Ivezi{\'c}}, {Kahn}, {Tyson}, {Abel}, {Acosta}, {Allsman}, {Alonso}, {AlSayyad}, {Anderson}, {Andrew}, et~al.]{ref_lsst}
{Ivezi{\'c}}, {\v Z}., {Kahn}, S.~M., {Tyson}, J.~A., {Abel}, B., {Acosta}, E., {Allsman}, R., {Alonso}, D., {AlSayyad}, Y., {Anderson}, S.~F., {Andrew}, J., et~al.
\newblock {LSST: From Science Drivers to Reference Design and Anticipated Data Products}.
\newblock \emph{The Astrophysical Journal}, 873:\penalty0 111, 2019.
\newblock \doi{10.3847/1538-4357/ab042c}.

\bibitem[Koblischke \& Bovy(2024)Koblischke and Bovy]{ref_koblischke2024}
Koblischke, N. and Bovy, J.
\newblock {SpectraFM: Tuning into Stellar Foundation Models}.
\newblock \emph{ArXiv e-prints}, 2024.
\newblock \doi{10.48550/arXiv.2411.04750}.

\bibitem[Leung \& Bovy(2023)Leung and Bovy]{ref_leung2023}
Leung, H.~W. and Bovy, J.
\newblock {Towards an astronomical foundation model for stars with a transformer-based model}.
\newblock \emph{Monthly Notices of the Royal Astronomical Society}, 527\penalty0 (1):\penalty0 1494--1520, 2023.
\newblock ISSN 0035-8711.
\newblock \doi{10.1093/mnras/stad3015}.

\bibitem[Loshchilov \& Hutter(2017)Loshchilov and Hutter]{ref_loshchilov2017}
Loshchilov, I. and Hutter, F.
\newblock {Decoupled Weight Decay Regularization}.
\newblock \emph{ArXiv e-prints}, 2017.
\newblock \doi{10.48550/arXiv.1711.05101}.

\bibitem[McInnes et~al.(2018)McInnes, Healy, and Melville]{ref_mcinnes2018}
McInnes, L., Healy, J., and Melville, J.
\newblock {UMAP: Uniform Manifold Approximation and Projection for Dimension Reduction}.
\newblock \emph{ArXiv e-prints}, 2018.
\newblock \doi{10.48550/arXiv.1802.03426}.

\bibitem[Mousavi et~al.(2025)Mousavi, Maimon, Moumen, Petermann, Shi, Wu, Yang, Kuznetsova, Ploujnikov, Marxer, Ramabhadran, Elizalde, Lugosch, Li, Subakan, Woodland, Kim, Lee, Watanabe, Adi, and Ravanelli]{ref_mousavi2025}
Mousavi, P., Maimon, G., Moumen, A., Petermann, D., Shi, J., Wu, H., Yang, H., Kuznetsova, A., Ploujnikov, A., Marxer, R., Ramabhadran, B., Elizalde, B., Lugosch, L., Li, J., Subakan, C., Woodland, P., Kim, M., Lee, H.-y., Watanabe, S., Adi, Y., and Ravanelli, M.
\newblock {Discrete Audio Tokens: More Than a Survey!}
\newblock \emph{ArXiv e-prints}, 2025.
\newblock \doi{10.48550/arXiv.2506.10274}.

\bibitem[Parker et~al.(2025)Parker, Lanusse, Shen, Liu, Hehir, Sarra, Meyer, Bowles, Wagner-Carena, Qu, Golkar, Bietti, Bourfoune, Casserau, Cornette, Hirashima, Krawezik, Ohana, Lourie, McCabe, Morel, Mukhopadhyay, Pettee, Blancard, Cho, Cranmer, and Ho]{ref_parker2025}
Parker, L., Lanusse, F., Shen, J., Liu, O., Hehir, T., Sarra, L., Meyer, L., Bowles, M., Wagner-Carena, S., Qu, H., Golkar, S., Bietti, A., Bourfoune, H., Casserau, N., Cornette, P., Hirashima, K., Krawezik, G., Ohana, R., Lourie, N., McCabe, M., Morel, R., Mukhopadhyay, P., Pettee, M., Blancard, B. R.-S., Cho, K., Cranmer, M., and Ho, S.
\newblock {AION-1: Omnimodal Foundation Model for Astronomical Sciences}.
\newblock \emph{ArXiv e-prints}, 2025.
\newblock \doi{10.48550/arXiv.2510.17960}.

\bibitem[Pearson(1901)]{ref_pearson1901}
Pearson, K.
\newblock {LIII. On lines and planes of closest fit to systems of points in space}.
\newblock \emph{London, Edinburgh, and Dublin Philosophical Magazine and Journal of Science}, 2\penalty0 (11):\penalty0 559--572, 1901.
\newblock ISSN 1941-5982.
\newblock \doi{10.1080/14786440109462720}.

\bibitem[Radford et~al.(2019)Radford, Wu, Child, Luan, A., and Sutskever]{ref_radford2019}
Radford, A., Wu, J., Child, R., Luan, D., A., D., and Sutskever, I.
\newblock Language models are unsupervised multitask learners.
\newblock \emph{OpenAI Whitepaper}, 2019.
\newblock URL \url{https://cdn.openai.com/better-language-models/language_models_are_unsupervised_multitask_learners.pdf}.

\bibitem[Raichoor et~al.(2023)Raichoor, Moustakas, Newman, Karim, Ahlen, Alam, Bailey, Brooks, Dawson, de~la Macorra, de~Mattia, Dey, Dey, Dhungana, Eftekharzadeh, Eisenstein, Fanning, Font-Ribera, Garc{\ifmmode\acute{\imath}\else\'{\i}\fi}a-Bellido, Gazta{\ifmmode\tilde{n}\else\~{n}\fi}aga, Gontcho, Guy, Honscheid, Ishak, Kehoe, Kisner, Kremin, Lan, Landriau, Le~Guillou, Levi, Magneville, Manera, Martini, Meisner, Myers, Nie, Palanque-Delabrouille, Percival, Poppett, Prada, Ross, Ruhlmann-Kleider, Sabiu, Schlafly, Schlegel, Tarl{\ifmmode\acute{e}\else\'{e}\fi}, Weaver, Y{\ifmmode\grave{e}\else\`{e}\fi}che, Zhou, Zhou, and Zou]{ref_raichoor2023}
Raichoor, A., Moustakas, J., Newman, J.~A., Karim, T., Ahlen, S., Alam, S., Bailey, S., Brooks, D., Dawson, K., de~la Macorra, A., de~Mattia, A., Dey, A., Dey, B., Dhungana, G., Eftekharzadeh, S., Eisenstein, D.~J., Fanning, K., Font-Ribera, A., Garc{\ifmmode\acute{\imath}\else\'{\i}\fi}a-Bellido, J., Gazta{\ifmmode\tilde{n}\else\~{n}\fi}aga, E., Gontcho, S. G.~A., Guy, J., Honscheid, K., Ishak, M., Kehoe, R., Kisner, T., Kremin, A., Lan, T.-W., Landriau, M., Le~Guillou, L., Levi, M.~E., Magneville, C., Manera, M., Martini, P., Meisner, A.~M., Myers, A.~D., Nie, J., Palanque-Delabrouille, N., Percival, W.~J., Poppett, C., Prada, F., Ross, A.~J., Ruhlmann-Kleider, V., Sabiu, C.~G., Schlafly, E.~F., Schlegel, D., Tarl{\ifmmode\acute{e}\else\'{e}\fi}, G., Weaver, B.~A., Y{\ifmmode\grave{e}\else\`{e}\fi}che, C., Zhou, R., Zhou, Z., and Zou, H.
\newblock {Target Selection and Validation of DESI Emission Line Galaxies}.
\newblock \emph{Astronomical Journal}, 165\penalty0 (3):\penalty0 126, 2023.
\newblock ISSN 1538-3881.
\newblock \doi{10.3847/1538-3881/acb213}.

\bibitem[Sanjaripour et~al.(2024)Sanjaripour, Hemmati, Mobasher, Canalizo, Barish, Shivaei, Coil, Chartab, Jafariyazani, Reddy, and Azadi]{sanjaripour2024}
Sanjaripour, S., Hemmati, S., Mobasher, B., Canalizo, G., Barish, B.~C., Shivaei, I., Coil, A.~L., Chartab, N., Jafariyazani, M., Reddy, N.~A., and Azadi, M.
\newblock The application of manifold learning to a selection of different galaxy populations and scaling relation analysis.
\newblock \emph{The Astrophysical Journal}, 977\penalty0 (2):\penalty0 202, dec 2024.
\newblock \doi{10.3847/1538-4357/ad90ba}.
\newblock URL \url{https://doi.org/10.3847/1538-4357/ad90ba}.

\bibitem[Sanjaripour et~al.(2025)Sanjaripour, Aravindan, Canalizo, Hemmati, Mobasher, Coil, and Barish]{sanjaripour2025}
Sanjaripour, S., Aravindan, A., Canalizo, G., Hemmati, S., Mobasher, B., Coil, A.~L., and Barish, B.~C.
\newblock Selection of dwarf galaxies hosting active galactic nuclei: A measure of bias and contamination using unsupervised machine learning techniques.
\newblock \emph{The Astrophysical Journal}, 992\penalty0 (1):\penalty0 138, oct 2025.
\newblock \doi{10.3847/1538-4357/ae0326}.
\newblock URL \url{https://doi.org/10.3847/1538-4357/ae0326}.

\bibitem[Smith \& Geach(2023)Smith and Geach]{ref_smith2023}
Smith, M.~J. and Geach, J.~E.
\newblock {Astronomia ex machina: a history, primer and outlook on neural networks in astronomy}.
\newblock \emph{R. Soc. Open Sci.}, 10\penalty0 (5):\penalty0 221454, 2023.
\newblock ISSN 2054-5703.
\newblock \doi{10.1098/rsos.221454}.

\bibitem[Smith et~al.(2024)Smith, Roberts, Angeloudi, and Huertas-Company]{ref_smith2024}
Smith, M.~J., Roberts, R.~J., Angeloudi, E., and Huertas-Company, M.
\newblock {AstroPT: Scaling Large Observation Models for Astronomy}.
\newblock \emph{ArXiv e-prints}, 2024.
\newblock \doi{10.48550/arXiv.2405.14930}.

\bibitem[Strubell et~al.(2019)Strubell, Ganesh, and Mccallum]{ref_strubell2019}
Strubell, E., Ganesh, A., and Mccallum, A.
\newblock {Energy and Policy Considerations for Deep Learning in NLP}.
\newblock \emph{ACL Anthology}, pp.\  3645--3650, 2019.
\newblock \doi{10.18653/v1/P19-1355}.

\bibitem[{The Multimodal Universe Collaboration}(2024)]{ref_mmu2024}
{The Multimodal Universe Collaboration}.
\newblock {The Multimodal Universe: Enabling Large-Scale Machine Learning with 100 TB of Astronomical Scientific Data}.
\newblock \emph{Advances in Neural Information Processing Systems}, 37:\penalty0 57841--57913, 2024.

\bibitem[Tschannen et~al.(2024)Tschannen, Pinto, and Kolesnikov]{ref_tschannen2024}
Tschannen, M., Pinto, A.~S., and Kolesnikov, A.
\newblock {JetFormer: An Autoregressive Generative Model of Raw Images and Text}.
\newblock \emph{ArXiv e-prints}, 2024.
\newblock \doi{10.48550/arXiv.2411.19722}.

\bibitem[van~den Oord et~al.(2017)van~den Oord, Vinyals, and Kavukcuoglu]{Oord_2017}
van~den Oord, A., Vinyals, O., and Kavukcuoglu, K.
\newblock Neural discrete representation learning.
\newblock In \emph{Neural Information Processing Systems}, 2017.
\newblock URL \url{https://api.semanticscholar.org/CorpusID:20282961}.

\bibitem[Vaswani et~al.(2017)Vaswani, Shazeer, Parmar, Uszkoreit, Jones, Gomez, Kaiser, and Polosukhin]{ref_vaswani2017}
Vaswani, A., Shazeer, N., Parmar, N., Uszkoreit, J., Jones, L., Gomez, A.~N., Kaiser, L., and Polosukhin, I.
\newblock {Attention Is All You Need}.
\newblock \emph{arXiv}, 2017.
\newblock \doi{10.48550/arXiv.1706.03762}.

\bibitem[Walmsley et~al.(2023)Walmsley, G{\ifmmode\acute{e}\else\'{e}\fi}ron, Kruk, Scaife, Lintott, Masters, Dawson, Dickinson, Fortson, Garland, et~al.]{ref_walmsley2023}
Walmsley, M., G{\ifmmode\acute{e}\else\'{e}\fi}ron, T., Kruk, S., Scaife, A. M.~M., Lintott, C., Masters, K.~L., Dawson, J.~M., Dickinson, H., Fortson, L., Garland, I.~L., et~al.
\newblock {Galaxy Zoo DESI: Detailed morphology measurements for 8.7M galaxies in the DESI Legacy Imaging Surveys}.
\newblock \emph{Monthly Notices of the Royal Astronomical Society}, 526\penalty0 (3):\penalty0 4768--4786, 2023.
\newblock ISSN 0035-8711.
\newblock \doi{10.1093/mnras/stad2919}.

\bibitem[Yu et~al.(2021)Yu, Li, Koh, Zhang, Pang, Qin, Ku, Xu, Baldridge, and Wu]{yu_2021}
Yu, J., Li, X., Koh, J.~Y., Zhang, H., Pang, R., Qin, J., Ku, A., Xu, Y., Baldridge, J., and Wu, Y.
\newblock Vector-quantized image modeling with improved vqgan.
\newblock \emph{arXiv preprint arXiv:2110.04627}, 2021.

\end{thebibliography}

\newpage
\appendix
\onecolumn
\renewcommand{\thefigure}{\Alph{section}.\arabic{figure}}
\renewcommand{\thetable}{\Alph{section}.\arabic{table}}
\setcounter{figure}{0}
\setcounter{table}{0}

\section{Embedding Structure via PCA and UMAP}
\label{app:embeddings}

\begin{figure*}[htbp]
    \centering
    \includegraphics[width=0.8\textwidth]{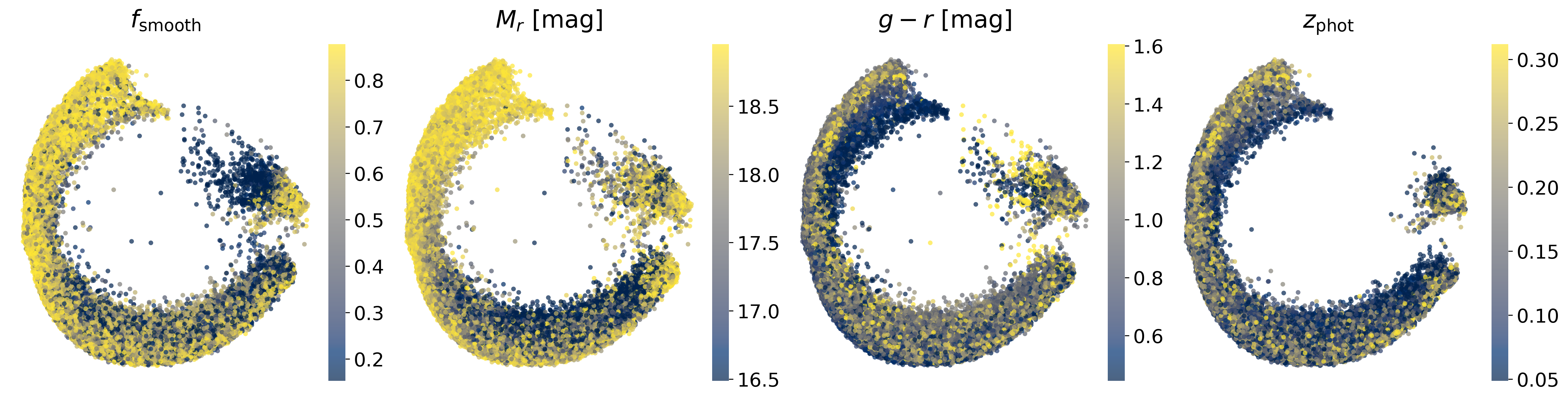}\\[4pt]
    \includegraphics[width=0.8\textwidth]{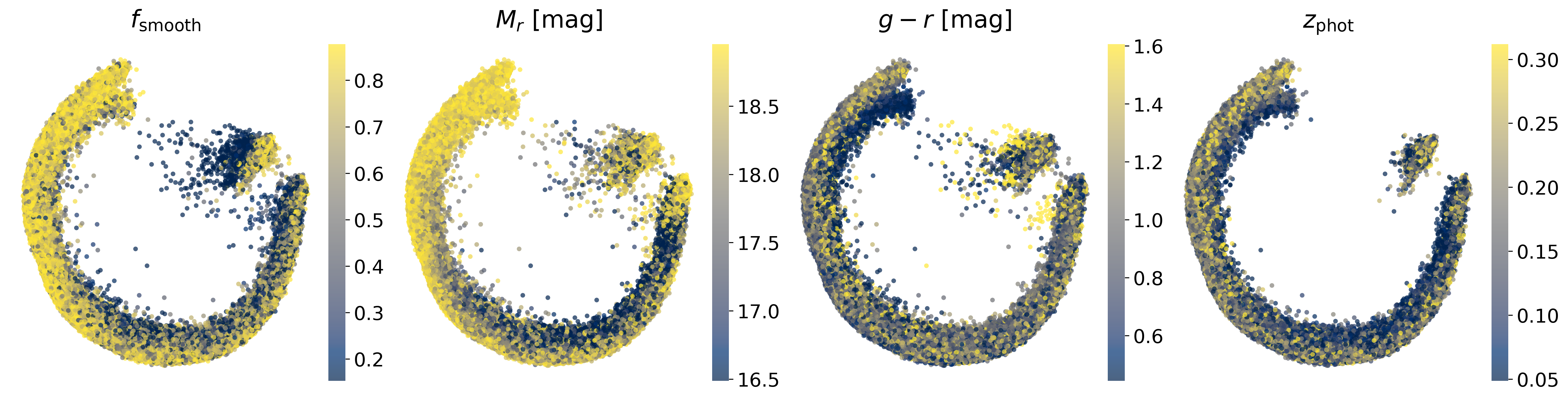}\\[4pt]
    \includegraphics[width=0.8\textwidth]{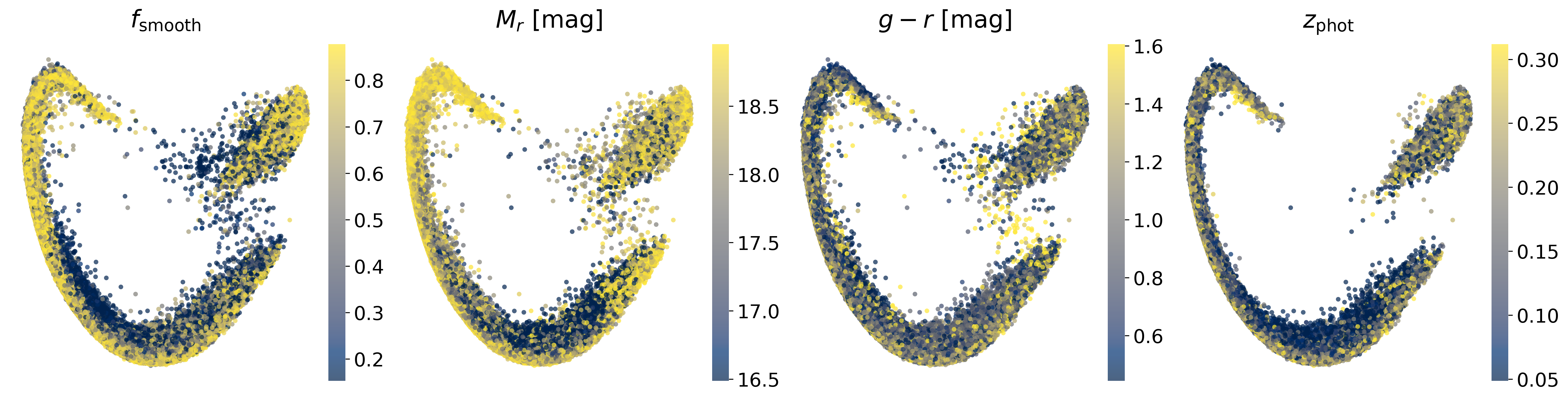}\\[4pt]
    \includegraphics[width=0.8\textwidth]{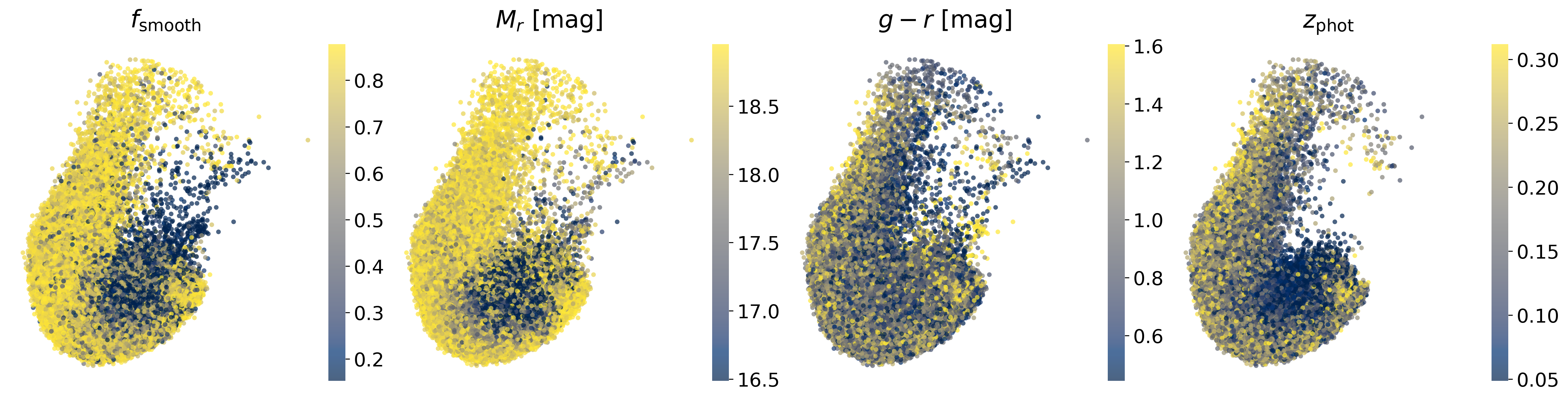}
    \caption{Comparison of galaxy representations obtained using different tokenization and embedding strategies, all based on the same transformer backbone (AstroPT). From top to bottom row: (a) Affine tokenizer, (b) AIM tokenizer, (c) JetFormer tokenizer, and (d) VQ-VAE tokenizer. Each column shows the embeddings color-coded by different galaxy properties, from left to right: smoothness fraction, DESI $r$-band magnitude, DESI $g-r$ photometric color, and photometric redshift. In each panel, the learned latent embeddings are projected onto two dimensions using principal component analysis (PCA).}
    \label{fig:PCA}
\end{figure*}

\begin{figure*}[t!]
    \centering
    \includegraphics[width=0.8\textwidth]{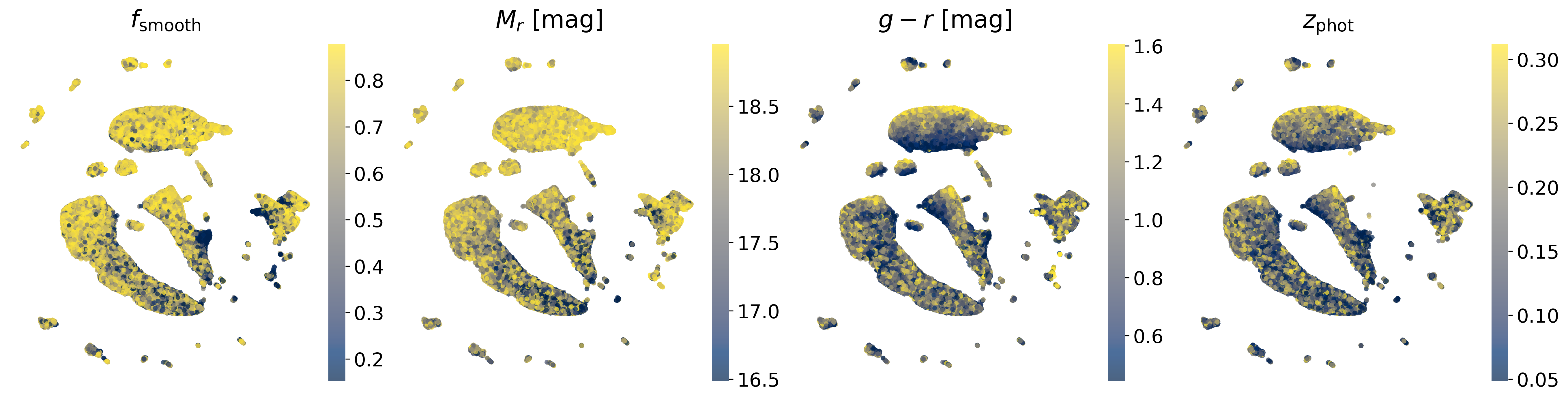}\\[4pt]
    \includegraphics[width=0.8\textwidth]{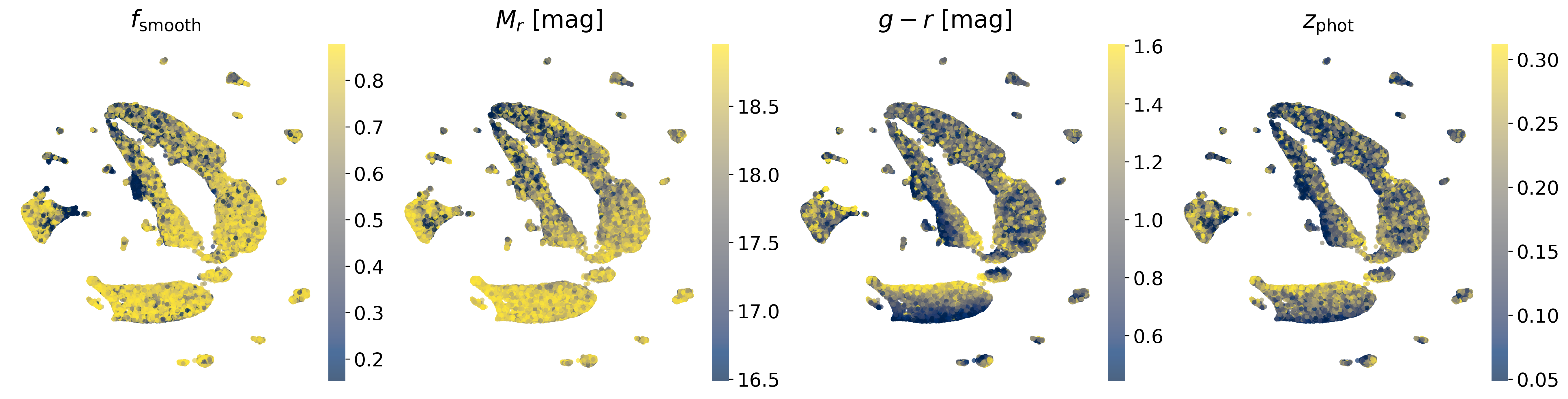}\\[4pt]
    \includegraphics[width=0.8\textwidth]{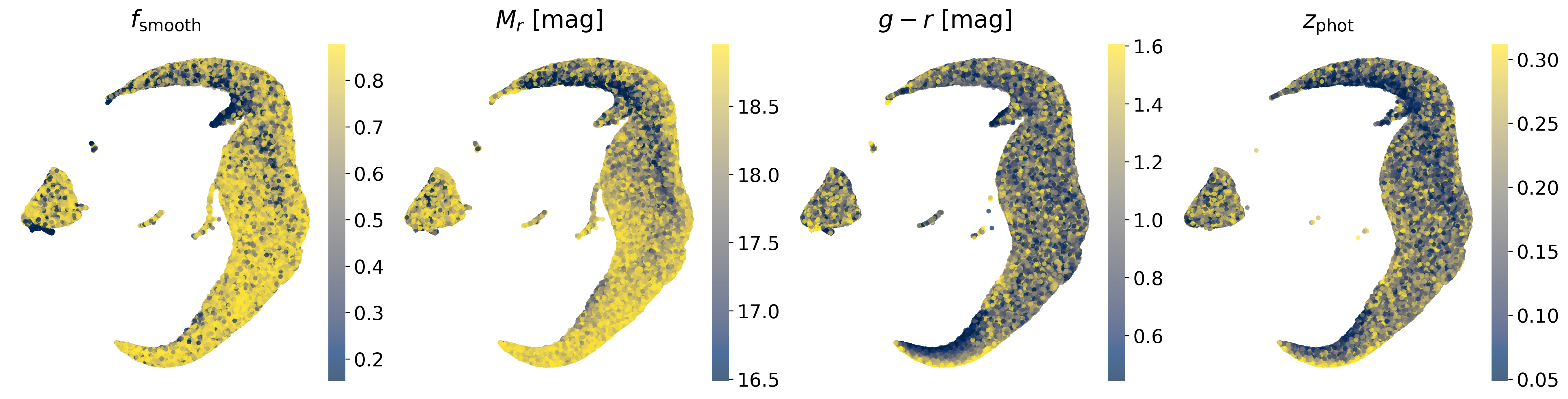}\\[4pt]
    \includegraphics[width=0.8\textwidth]{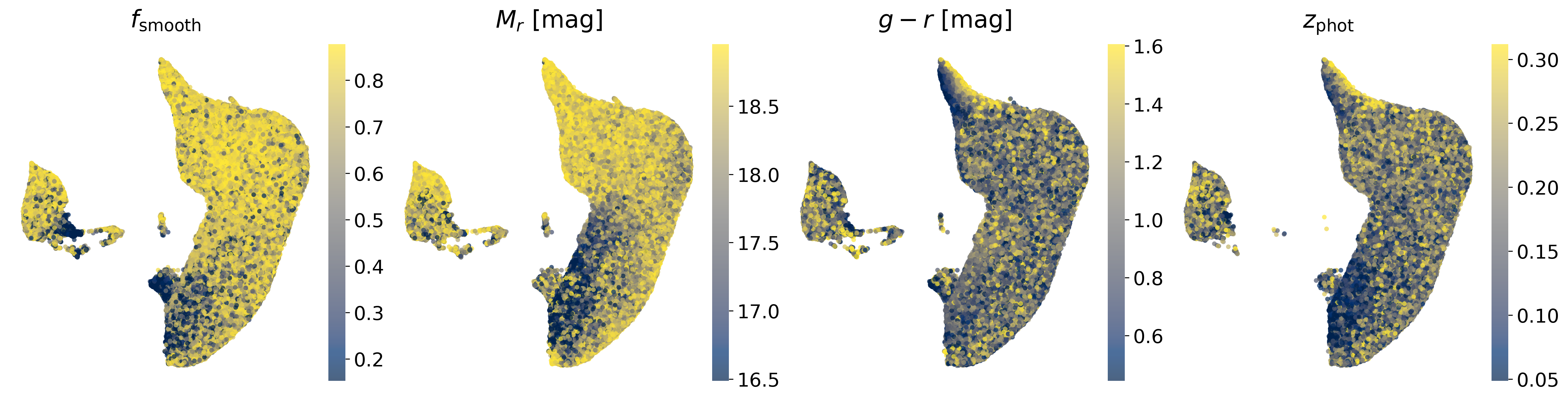}
    \caption{Comparison of galaxy representations obtained using different tokenization and embedding strategies, all based on the same transformer backbone (AstroPT). From top to bottom row: (a) Affine tokenizer, (b) AIM tokenizer, (c) JetFormer tokenizer, and (d) VQ-VAE tokenizer. Each column shows the embeddings color-coded by different galaxy properties, from left to right: smoothness fraction, DESI $r$-band magnitude, DESI $g-r$ photometric color, and photometric redshift. In each panel, the learned latent embeddings are projected onto two dimensions using UMAP.}
    \label{fig:UMAP}
\end{figure*}

We project the 768-dimensional embeddings onto two dimensions using PCA and UMAP analysis \citep{ref_pearson1901,ref_mcinnes2018}.
The resulting projections are colour-coded by photometric redshift, $g-r$ colour, $r$-band magnitude, and smoothness fraction as measured by Galaxy~Zoo Citizen Scientist answers.
As shown in Figures~\ref{fig:PCA} and~\ref{fig:UMAP}, both projections reveal coherent gradients in physical parameters, most notably a strong correlation between $g-r$ colour and photometric redshift.
This behaviour reflects the intrinsic coupling between galaxy spectral properties and distance, and indicates that the learned embeddings encode physically meaningful information directly from imaging data.

Beyond this colour--redshift coupling, the embeddings organise along a second, largely distinct direction traced by the smoothness fraction and $r$-band magnitude, with smoother (early-type) galaxies tending to occupy the brighter end of the distribution.
The latent space therefore encodes at least two semantically separate axes of variation---one governing colour and distance, the other morphology and luminosity---recovering well-known scaling relations directly from imaging.

The projected spaces also exhibit distinct structural patterns across the different tokenisation methods.
Under PCA, Affine and AIM collapse onto an open, horseshoe-shaped arc characteristic of a single dominant variance direction, JetFormer forms a more elongated crescent, and VQ-VAE produces a compact, filled lobe.
The same ordering is visible under UMAP: Affine and AIM fragment into an archipelago of many disconnected islands, whereas JetFormer and VQ-VAE retain a single connected manifold.
Crucially, the physical gradients persist across both projections and remain coherent even across the fragmented UMAP clusters, indicating that the global ordering in property space is intrinsic to the representation rather than an artefact of any one dimensionality-reduction method.

Consistent with the near-identical probe results, Affine and AIM produce visually similar (albeit transposed) embedding geometries.
We note, however, that this apparent transposition reflects only the sign and rotation ambiguity inherent to PCA eigenvectors and carries no physical meaning; together with their matching island structure under UMAP and their matching probe scores, it reinforces the hypothesis that the MLP tokeniser head contributes little beyond what a linear projection provides.

\section{Computational Cost and Carbon Footprint}
\label{app:carbon}
The training of deep learning models requires considerable energy, contributing to carbon emissions.
To counteract further emission from redundant retraining, we follow the recommendations of \citet{ref_strubell2019} and will make available our fully trained models and code upon deanonymization.
Affine and AIM were each trained for 3~hours on a single NVIDIA H100 80~GB GPU, corresponding to an estimated energy consumption of 2.1~kWh, while JetFormer required 8~hours of training on two NVIDIA H100 80~GB GPUs, resulting in a total estimated energy consumption of 11.2~kWh based on reported GPU TDP values. In comparison, VQ-VAE required approximately 20~hours of training on an NVIDIA Quadro GV100 GPU with 32~GB memory, corresponding to an estimated energy consumption of 5.2~kWh.

\begin{figure}[h]
  \centering
  \includegraphics[width=\textwidth]{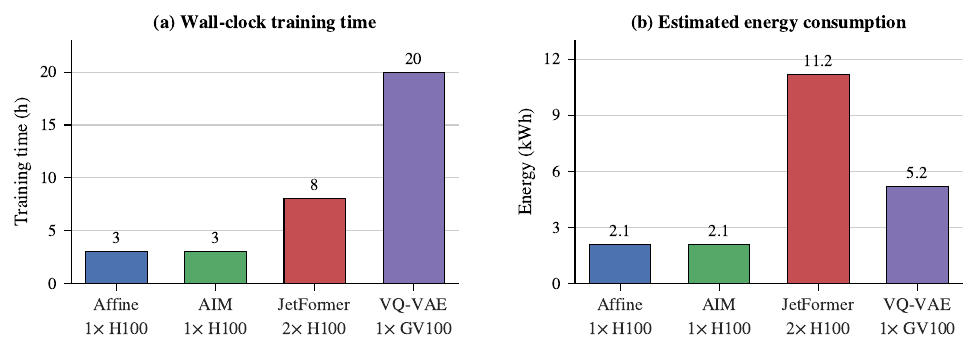}
  \caption{Training cost of the four tokenisation strategies on a shared
  AstroPT backbone. \textbf{(a)} Wall-clock training time and \textbf{(b)}
  estimated energy consumption, with the GPU configuration annotated beneath
  each bar. Affine and AIM were trained on a single NVIDIA H100 80~GB GPU,
  JetFormer on two NVIDIA H100 80~GB GPUs, and VQ-VAE on a single NVIDIA
  Quadro GV100 32~GB GPU; energy is estimated from reported GPU TDP values.
  Because the runtimes were measured on different hardware, energy provides a
  fairer cross-tokeniser comparison than wall-clock time: VQ-VAE trains
  longest yet, on the lower-power GV100, consumes less than half the energy of
  JetFormer's dual-H100 configuration, while Affine and AIM are the cheapest
  on both axes.}
  \label{fig:tokenizer_cost}
\end{figure}

\end{document}